\documentclass[final,5p,times,float,twocolumn]{elsarticle}

\usepackage{amssymb}
\usepackage{amsthm}
\usepackage{amsmath}

\usepackage{graphicx}
\usepackage{dcolumn}
\usepackage{bm}
\usepackage{multirow}
\usepackage{tabularx}
\usepackage{caption}
\usepackage[colorlinks=true,linkcolor=blue,citecolor=blue,urlcolor=blue]{hyperref}
\usepackage{booktabs}
\usepackage{comment} 
\usepackage[T5,T1]{fontenc} 
\DeclareUnicodeCharacter{2212}{-}

\journal{Physics Letters B}

\usepackage{xspace}
\usepackage[normalem]{ulem}   
\newcommand{\pieter}{\textcolor{red}}

\newcommand{\ts}{\textsuperscript}

\begin{document}

\begin{frontmatter}



\title{Spectroscopy of deeply bound orbitals in neutron-rich Ca isotopes}


%

\author[alanzhou,ahku]{P.~J.~Li\corref{correspondingauthor1}}
\cortext[correspondingauthor1]{Corresponding author}\ead{lipengjie@impcas.ac.cn}
\author[ahku]{J.~Lee\corref{correspondingauthor2}}
\cortext[correspondingauthor2]{Corresponding author}\ead{jleehc@hku.hk}
\author[ariken]{P.~Doornenbal}
\author[ahku,ariken,ayork]{S.~Chen}
\author[alanzhou]{S.~Wang}
\author[atudarmstadt,acea,ariken]{A.~Obertelli} 
\author[ariken]{Y.~Chazono} 
\author[atriumf,amcgill]{J.~D.~Holt} 
\author[atriumf,anccs,apd]{B.~S.~Hu} 
\author[akyushu,arcnp]{K.~Ogata} 
\author[ajaea,acns]{Y.~Utsuno} 
\author[ajaea]{K.~Yoshida} 
\author[acaen]{N.~L.~Achouri}
\author[ariken]{H.~Baba} 
\author[ariken]{F.~Browne}
\author[acea]{D.~Calvet} 
\author[acea]{F.~Ch\^ateau} 
\author[ariken]{N.~Chiga} 
\author[acea]{A.~Corsi}  
\author[ariken]{M.~L.~Cort\'es} 
\author[acea]{A.~Delbart} 
\author[acea]{J-M.~Gheller} 
\author[acea]{A.~Giganon}                           
\author[acea]{A.~Gillibert} 
\author[acea]{C.~Hilaire} 
\author[ariken]{T.~Isobe}  
\author[atohoku]{T.~Kobayashi}
\author[ariken,acns]{Y.~Kubota} 
\author[acea]{V.~Lapoux} 
\author[aklbtme,atudarmstadt]{H.~N.~Liu}       
\author[ariken]{T.~Motobayashi}                     
\author[aijclab,ariken]{I.~Murray} 
\author[ariken]{H.~Otsu} 
\author[ariken]{V.~Panin}
\author[acea,abro]{N.~Paul}                        
\author[ariken,ajaver,aunal]{W.~Rodriguez} 
\author[ariken,aut]{H.~Sakurai} 
\author[ariken]{M.~Sasano} 
\author[ariken]{D.~Steppenbeck}
\author[acns,aatomki,aibs]{L.~Stuhl}              
\author[acea,atudarmstadt]{Y.~L.~Sun}            
\author[arikkyo,ariken]{Y.~Togano}  
\author[ariken]{T.~Uesaka} 
\author[aut,ariken]{K.~Wimmer}                     
\author[ariken]{K.~Yoneda}  
\author[akth]{O.~Aktas}
\author[atudarmstadt,agsi]{T.~Aumann}              
\author[agsi]{K.~Boretzky}                          
\author[atudarmstadt,agsi,ariken]{C.~Caesar}      
\author[ainst]{L.~X.~Chung}  
\author[aijclab]{F.~Flavigny} 
\author[aijclab]{S.~Franchoo} 
\author[azagreb,atudarmstadt,ariken]{I.~Gasparic} 
\author[akoeln]{R.-B.~Gerst}
\author[acaen]{J.~Gibelin}
\author[aewha,aibs]{K.~I.~Hahn} 
\author[atudarmstadt]{J.~Kahlbow}                   
\author[aewha,ariken,aibs]{D.~Kim} 
\author[aut]{T.~Koiwai}                             
\author[atitech]{Y.~Kondo}                          
\author[agsi]{D.~K\"orper}                          
\author[atudarmstadt,agsi]{P.~Koseoglou}
\author[atudarmstadt]{C.~Lehr}                      
\author[ainst,avarns]{B.~D.~Linh} 
\author[ahku]{T.~Lokotko}
\author[aijclab]{M.~MacCormick}
\author[atudarmstadt,amsu]{K.~Miki}                
\author[akoeln]{K.~Moschner}
\author[atitech]{T.~Nakamura}                       
\author[aewha,aibs]{S.~Y.~Park}  
\author[atudarmstadt,agsi]{D.~Rossi}                     
\author[aoslo]{E.~Sahin} 
\author[atudarmstadt]{F.~Schindler}                 
\author[agsi]{H.~Simon}                             
\author[atudarmstadt,aelinp]{P-A.~S\"oderstr\"om} 
\author[aatomki]{D.~Sohler}  
\author[atitech]{S.~Takeuchi} 
\author[atudarmstadt,agsi]{H.~Toernqvist}         
\author[atudarmstadt]{J.~Tscheuschner}             
\author[amadrid]{V.~Vaquero} 
\author[atudarmstadt]{V.~Wagner}                    
\author[atudarmstadt]{V.~Werner}
\author[ahku]{X.~Xu} 
\author[atitech]{H.~Yamada}                         
\author[alanzhou]{D.~Yan} 
\author[ariken]{Z.~Yang}  
\author[atitech]{M.~Yasuda}                         
\author[atudarmstadt]{L.~Zanetti}                   

\affiliation[alanzhou]{
    organization={CAS Key Laboratory of High Precision Nuclear Spectroscopy, Institute of Modern Physics, Chinese Academy of Sciences},
    addressline={Lanzhou 730000},
    country={China}
}
\affiliation[ahku]{
    organization={Department of Physics, The University of Hong Kong}, 
    addressline={Pokfulam, Hong Kong},
    country={China}
}
\affiliation[ariken]{
    organization={RIKEN Nishina Center}, 
    addressline={2-1 Hirosawa, Wako, Saitama 351-0198},
    country={Japan}
}
\affiliation[ayork]{
    organization={School of Physics, Engineering and Technology, University of York}, 
    addressline={Heslington, York, YO10 5DD},
    country={UK}
 }
\affiliation[atudarmstadt]{
    organization={Institut f\"ur Kernphysik, Technische Universit\"at Darmstadt}, 
    addressline={64289 Darmstadt}, 
    country={Germany}
}
\affiliation[acea]{
    organization={IRFU, CEA, Universit\'e Paris-Saclay}, 
    addressline={F-91191 Gif-sur-Yvette},
    country={France}
}
\affiliation[atriumf]{
    organization={TRIUMF}, 
    addressline={4004 Wesbrook Mall, Vancouver, British Columbia V6T 2A3},
    country={Canada}
}
\affiliation[amcgill]{
    organization={Department of Physics, McGill University}, 
    addressline={Montréal, Quebec City H3A 2T8},
    country={Canada}
}
\affiliation[anccs]{
    organization={National Center for Computational Sciences},
    addressline={Oak Ridge National Laboratory, Oak Ridge, Tennessee 37831},
    country={USA}
}
\affiliation[apd]{
    organization={Physics Division},
    addressline={Oak Ridge National Laboratory, Oak Ridge, Tennessee 37831},
    country={USA}
}
\affiliation[akyushu]{
    organization={Department of Physics, Kyushu University}, 
    addressline={Fukuoka 819-0395},
    country={Japan}
}
\affiliation[arcnp]{
    organization={Research Center for Nuclear Physics (RCNP), Osaka University}, 
    addressline={Ibaraki},
    country={Japan}
}
\affiliation[ajaea]{
    organization={Advanced Science Research Center, Japan Atomic Energy Agency, Tokai}, 
    addressline={Ibaraki 319-1195},
    country={Japan}
}
\affiliation[acns]{
    organization={Center for Nuclear Study, University of Tokyo}, 
    addressline={RIKEN campus, Wako, Saitama 351-0198}, 
    country={Japan}
}
\affiliation[acaen]{
    organization={Université de Caen Normandie, ENSICAEN, CNRS/IN2P3, LPC Caen UMR6534}, 
    addressline={F-14000 Caen}, 
    country={France}
}
\affiliation[atohoku]{
    organization={Department of Physics, Tohoku University}, 
    addressline={Sendai 980-8578}, 
    country={Japan}
}
\affiliation[aklbtme]{
    organization={Key Laboratory of Beam Technology of Ministry of Education}, 
    addressline={College of Nuclear Science and Technology, Beijing Normal University,
Beijing 100875}, 
    country={China}
}
\affiliation[akth]{
    organization={Department of Physics, Royal Institute of Technology}, 
    addressline={SE-10691 Stockholm}, 
    country={Sweden}
}
\affiliation[aijclab]{
    organization={Universit\'e Paris-Saclay, CNRS/IN2P3, IJCLab}, 
    addressline={F-91405 Orsay cedex}, 
    country={France}
}
\affiliation[abro]{
    organization={Laboratoire Kastler Brossel, Sorbonne Universit\'e, CNRS, ENS, PSL Research University, Coll\`ege de France}, 
    addressline={Case 74, 4 Place Jussieu, 75005 Paris}, 
    country={France}
}
\affiliation[ajaver]{
    organization={Pontificia Universidad Javeriana, Facultad de Ciencias, Departamento de F\'isica},
    addressline={Bogot\'a}, 
    country={Colombia}
}
\affiliation[aunal]{
    organization={Universidad Nacional de Colombia, Sede Bogot\'a, Facultad de Ciencias, Departamento de Física}, 
    addressline={Bogot\'a 111321}, 
    country={Colombia}
}
\affiliation[aut]{
    organization={Department of Physics, University of Tokyo}, 
    addressline={7-3-1 Hongo, Bunkyo, Tokyo 113-0033},
    country={Japan}
}
\affiliation[aatomki]{
    organization={Institute for Nuclear Research, Atomki}, 
    addressline={P.O. Box 51, Debrecen H-4001},
    country={Hungary}
}
\affiliation[aibs]{
    organization={Institute for Basic Science}, 
    addressline={Daejeon 34126}, 
    country={Korea}
}
\affiliation[arikkyo]{
    organization={Department of Physics, Rikkyo University}, 
    addressline={3-34-1 Nishi-Ikebukuro, Toshima, Tokyo 172-8501}, 
    country={Japan}
}
\affiliation[agsi]{
    organization={GSI Helmholtzzentrum f\"ur Schwerionenforschung GmbH}, 
    addressline={Planckstr. 1, 64291 Darmstadt}, 
    country={Germany}
}
\affiliation[ainst]{
    organization={Institute for Nuclear Science \& Technology, VINATOM}, 
    addressline={179 Hoang Quoc Viet, Cau Giay, Hanoi},
    country={Vietnam}
}
\affiliation[azagreb]{
    organization={Ru{\dj}er Bo\v{s}kovi\'c Institute},
    addressline={Bijeni\v{c}ka cesta 54, 10000 Zagreb}, 
    country={Croatia}
}
\affiliation[akoeln]{
    organization={Institut f\"ur Kernphysik, Universit\"at zu K\"oln}, 
    addressline={D-50937 Cologne}, 
    country={Germany}
}
\affiliation[aewha]{
    organization={Ewha Womans University}, 
    addressline={Seoul 03760}, 
    country={Korea}
}
\affiliation[atitech]{
    organization={Department of Physics, Tokyo Institute of Technology}, 
    addressline={2-12-1 O-Okayama, Meguro, Tokyo, 152-8551}, 
    country={Japan}
}
\affiliation[avarns]{
    organization={Vietnam Agency for Radiation and Nuclear Safety}, 
    addressline={113 Tran Duy Hung, Cau Giay, Hanoi}, 
    country={Vietnam}
}
\affiliation[aoslo]{
    organization={Department of Physics, University of Oslo}, 
    addressline={N-0316 Oslo}, 
    country={Norway}
}
\affiliation[aelinp]{
    organization={Extreme Light Infrastructure-Nuclear Physics (ELI-NP)/Horia Hulubei National Institute for Physics and Nuclear Engineering (IFIN-HH)}, 
    addressline={Str. Reactorulu, M\u{a}gurele 077125}, 
    country={Romania}
}
\affiliation[amadrid]{
    organization={Instituto de Estructura de la Materia, CSIC}, 
    addressline={E-28006 Madrid}, 
    country={Spain}
}

\date{\today}

\begin{abstract}
The calcium isotopes are an ideal system to investigate the evolution of shell structure and magic numbers. 
Although the properties of surface nucleons in calcium have been well studied, probing the structure of deeply bound nucleons remains a challenge. 
Here, we report on the first measurement of unbound states in $^{53}$Ca and $^{55}$Ca, populated from \ts{54,56}Ca($p,pn$) reactions at a beam energy of around 216 MeV/nucleon at the RIKEN Radioactive Isotopes Beam Factory.
The resonance properties, partial cross sections, and momentum distributions of these unbound states were analyzed. 
Orbital angular momentum $l$ assignments were extracted from momentum distributions based on calculations using the distorted wave impulse approximation (DWIA) reaction model. The resonances at excitation energies of 5516(41)\,keV in $^{53}$Ca and 6000(250)\,keV in $^{55}$Ca indicate a significant $l$\, =\,3 component, providing the first experimental evidence for the $\nu 0f_{7/2}$ single-particle strength of unbound hole states in the neutron-rich Ca isotopes.
The observed excitation energies and cross-sections point towards extremely localized and well separated strength distributions, with some
fragmentation for the $\nu 0f_{7/2}$ orbital in $^{55}$Ca. These results are in good agreement with predictions from shell-model calculations using the effective GXPF1Bs interaction and \textit{ab initio} calculations and diverge markedly from the experimental distributions
in the nickel isotones at $Z=28$.
\end{abstract}
%



\begin{keyword}
unbound states, knockout reaction, single-particle strength, shell evolution
\end{keyword}

\end{frontmatter}




\section{Introduction}
Nuclear shell evolution towards the driplines poses a significant challenge in modern nuclear physics, characterized by the quenching or collapse of conventional magic numbers, and the reordering of single-particle orbitals, potentially giving rise to the emergence of new magic numbers and/or sub-shell closures.
The calcium isotopes feature a robust proton $Z = 20$ shell closure. Hence, the doubly magic isotopes $^{40}$Ca and $^{48}$Ca as well as experimental evidence of new neutron sub-shell closures at $N = 32$~\cite{wienholtz:2013:NATURE,enciu:2022:PRL} and 34~\cite{steppenbeck:2013:NATURE,michimasa:2018:PRL,chen:2019:PRL} provide a unique testing ground to study shell evolution. 

In the independent-particle shell model picture, nucleons occupy the single-particle orbitals and are bound in an attractive mean-field induced by the interactions among all nucleons. 
Ideally, occupation probabilities for the single-particle and single-hole states in the vicinity of a doubly magic nucleus should reflect the independent particle filling.
However, near the doubly magic nucleus $^{48}$Ca, a nearly 50$\%$ reduction in spectroscopic strength has been observed in the $0f_{7/2}$ neutron hole state populated from one-neutron knockout of $^{50}$Ca$\rightarrow$$^{49}$Ca, in contrast to the $^{48}$Ca$\rightarrow$$^{47}$Ca neutron knockout~\cite{crawford:2017:PRC,Macchiavelli:2022:EPJA}.
These results point to the fragmentation of angular momentum $l=3$ strength to higher-lying states and cannot be reproduced by any shell model calculations. Also for a systematic study of one neutron knockout reaction in $^{51-55}$Sc isotopes~\cite{schwertel:2012:EPJA} the observed spectroscopic strength of the $l = 3$ contribution was approximately half of the prediction from shell model calculations using the GXPF1A interaction~\cite{honma:2005:EPJA}, indicating a weakening of the $N = 28$ shell gap and a considerable fragmentation of neutron $0f_{7/2}$ spectroscopic strength in neutron-rich scandium isotopes\,($N = 30-34$).
In contrast, a recent experiment on the neutron-knockout from $^{52}$Ca employing the quasi-free ($p,pn$) scattering reaction has revealed a spectroscopic factor of $C^2S = 6.6(10)$ for the $\nu 0f_{7/2}$ hole state, suggesting a consistent shell structure in $^{48}$Ca and $^{52}$Ca~\cite{enciu:2022:PRL}. 

Given these experimental variations, further inspection regarding the evolution of the $\nu 0f_{7/2}$ single-particle orbital in the neutron-rich Ca isotopic chain has become prudent.
Early \textit{ab-initio} calculations using many-body perturbation theory and the low-momentum interaction $V_{\text{low}k}$ obtained by evolving a chiral N$^3$LO NN potential predict a decreasing and eventually vanishing gap in the effective single particle energies (ESPEs) between the $0f_{7/2}$ and $1p_{3/2}$ single particle energies (SPEs) for calcium isotopes beyond \ts{48}Ca for calculations in the $pfg_{9/2}$ model space~\cite{holt:2012:JPG,holt:2013:JPG,holt:2014:PRC}. Conversely, the phenomenological interactions GXPF1(A)~\cite{honma:2002:PRC,honma:2004:PRC,honma:2005:EPJA} and 
KB3G~\cite{poves:2001:NPA} yield a significant gap of $\approx$5 MeV for \ts{48}Ca, which, albeit decreasing, remains at approximately 3 MeV between \ts{54}Ca and \ts{60}Ca.
A similar trend is observed for the recent data driven, in-medium similarity renormalization group (IMSRG) based interaction UFPA-CA~\cite{magilligan:2021:PRC}. 


Here, we report on the first invariant-mass measurement of unbound states in $^{53,55}$Ca, populated using one-neutron quasi-free scattering reactions at relativistic energies. For the first time, the spectroscopic strength of the $\nu 0f_{7/2}$ neutron-hole states in $^{53}$Ca and $^{55}$Ca has been extracted by comparing experimental data to single-particle cross sections using DWIA calculations. 
In doing so, experimental knowledge on bound states of the isotopes $^{53,55}$Ca, acquired in recent one-neutron knockout studies~\cite{chen:2019:PRL,koiwai:2022:PLB} and an earlier $\beta$-decay investigation of \ts{53}K~\cite{perrot:2006:PRC}, is expanded significantly.
Our results show that the extracted spectroscopic strengths of the neutron $1p_{1/2}$, $1p_{3/2}$, and $0f_{5/2}$ hole states are consistent with shell model calculations based on the GXPF1Bs interaction~\cite{chen:2019:PRL} and the \textit{ab-initio} valence-space in-medium similarity renormalization group (VS-IMSRG)~\cite{stroberg:2019:ARNPS}.

No evidence for $\nu 0f_{7/2}$ strength below the neutron separation energy $S_n$ was found in earlier
experiments~\cite{chen:2019:PRL,koiwai:2022:PLB}, leading to the assumption that it is associated with states above the 
neutron decay threshold.
Our findings reveal that the $0f_{7/2}$ neutron-hole states in $^{53,55}$Ca maintain a nearly pure single-hole character at high excitation energies, demonstrating remarkable agreement with the theoretical predictions. Differences are found in the location of the 
$1p_{3/2}$ and $0f_{7/2}$ hole states, from which the propagation of the $N=28$ shell gap can be inferred. By completing the trend of spectroscopic strength distribution of the $pf$ SPEs  up to \ts{56}Ca, crucial insights into the evolution of shell structures in neutron-rich calcium isotopes is provided. 

\section{Experimental Setup}

The experiment was carried out at the Radioactive Isotope Beam Factory (RIBF) operated by the RIKEN Nishina Center and the Center for Nuclear Study (CNS), University of Tokyo.
Secondary beam cocktails were produced by fragmentation of a 345\,MeV/nucleon $^{70}$Zn primary beam that impinged on a 10-mm-thick $^{9}$Be target with a typical intensity of 240\,pnA.
The beams of interest were selected and transported to the secondary target using the BigRIPS two-stage fragment separator~\cite{kubo:2012:PTEP}.
The beam particles were identified in BigRIPS by their atomic number $Z$ and mass-to-charge ratio ($A/Q$) on an event-by-event basis, shown in Fig.~1(a) of Ref.~\cite{Browne:2021:PRL}, extracted by their energy loss in a set of plastic detectors and the B$\rho$-$\Delta$E-TOF method~\cite{Fukuda:2013:NIMB}.
The $^{54}$Ca and $^{56}$Ca beams bombarded the 151-mm-long liquid hydrogen (LH$_2$) target of the MINOS device~\cite{obertelli:2014:EPJA} with an average energy of $\sim$250\,MeV/nucleon to induce quasi-free $(p,pn)$ scattering reactions.
MINOS also included a 300-mm-long time projection chamber (TPC) surrounding the LH$_2$ target to reconstruct the trajectories of the recoil protons with a position resolution of $\sim$2\,mm ($\sigma$) in the beam direction~\cite{santamaria:2018:NIMA}.
Reaction residues were analyzed by the SAMURAI spectrometer~\cite{kobayashi:2013:NIMB} with a magnetic field set to 2.7\,T in the center. The particle identification from SAMURAI is shown in Fig 1(b) of Ref.~\cite{Browne:2021:PRL}, following a method similar as for the incoming beams.
Hydrogen targets simplify the knockout process by providing a straightforward proton probe, leading to improved access to deeply bound states compared to traditional nuclear targets like Be or C, as evidenced by ongoing efforts at facilities such as RIBF and GSI\cite{Aumann:2013:PRC,panin:2021:EPJA,Pohl:2023:PRL,Crawford:2020:APR,Petri:S467:R3B}.

Population to unbound states of $^{53,55}$Ca was followed by forward directed neutron emission. These approximately beam-velocity neutrons were detected by two large-acceptance plastic scintillator arrays, the NeuLAND demonstrator~\cite{boretzky:2021:NIMA} and NEBULA~\cite{kobayashi:2013:NIMB,nakamura:2016:NIMB}, located 11.8\,m and 14.85\,m downstream of the target, respectively.
The NeuLAND demonstrator consisted of 400 modules ($5\times5\times250$\,cm$^3$) in 8 layers.
The NEBULA array consisted of 120 neutron detection modules ($12\times12\times180$\,cm$^3$),
arranged in a two-wall configuration.
The neutron momenta were derived from their TOFs and the flight paths between the reaction vertex and the hit position on the neutron detectors.
Neutron detection efficiencies of $\varepsilon_{n}\approx37\%$ and $\varepsilon_{nn}\approx12\%$ for decay energies at 2\,MeV were simulated with the GEANT4~\cite{agostinelli:2003:NIMA} framework.

The de-excitation $\gamma$ rays emitted from fragments were detected by the DALI2$^+$ array~\cite{takeuchi:2014:NIMA,murray:2018:RAPR}, consisting of 226 NaI(Tl) crystals that surrounded the MINOS device. An add-back procedure combining hits within 15\,cm was applied during the analysis to increase the photopeak efficiency.
For a 2-MeV $\gamma$ ray emitted by particles moving at 60\% of the speed-of-light, the photopeak efficiency was $23\%$ with add-back, and the energy resolution after Doppler correction was $75$\,keV($\sigma$).
\section{Results}
\begin{figure}[htbp]
    \centering
    \includegraphics[width=\linewidth]{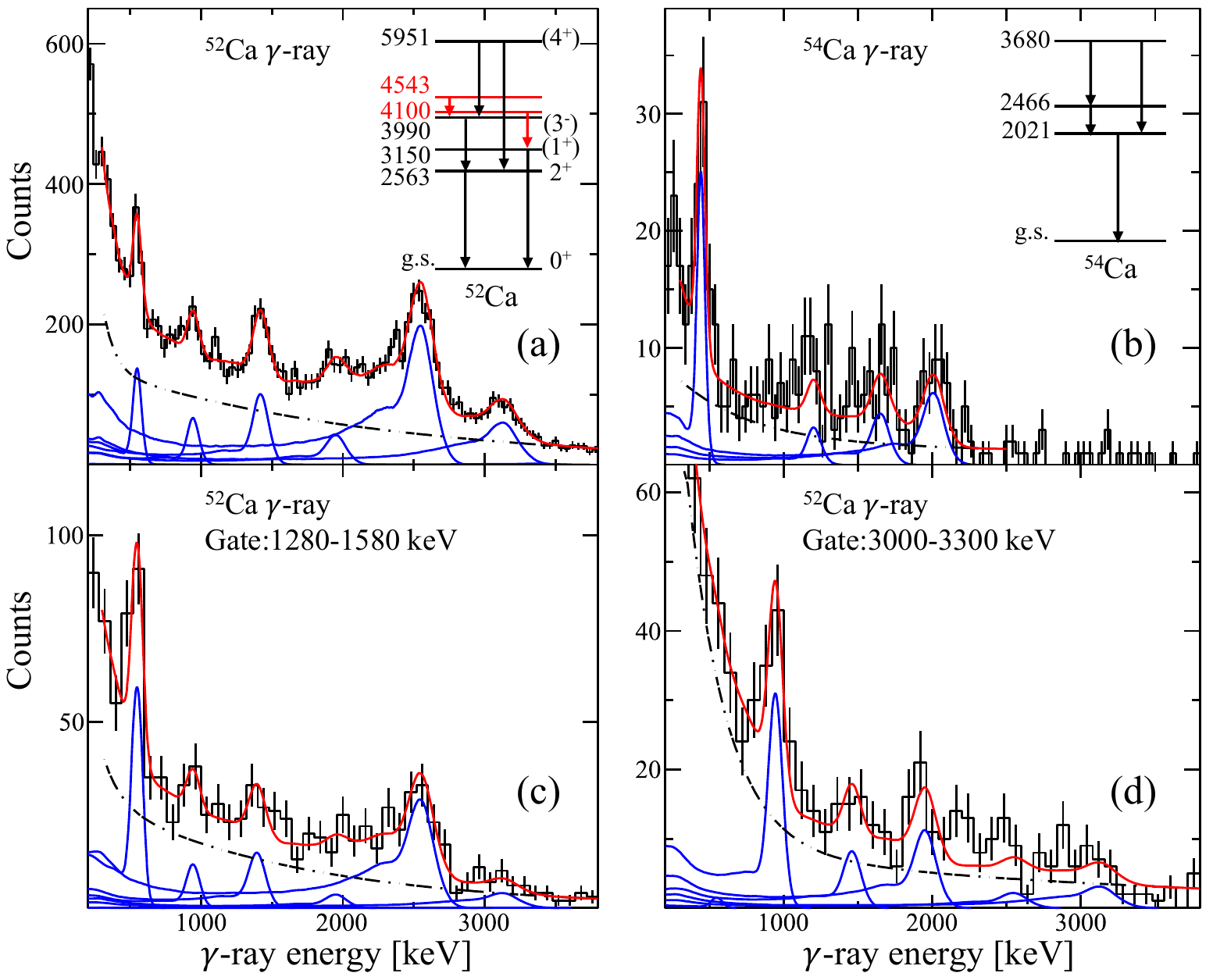}
    \caption{
    Doppler-corrected de-excitation $\gamma$-ray spectra of $^{52,54}$Ca fragments following neutron decays. (a) and (b) show the total spectra for $^{52}$Ca and $^{54}$Ca, respectively.
    The fit function (red line) comprises a double-exponential background (black dash dotted line) and simulated $\gamma$-ray response functions of DALI2$^+$ (blue lines). 
    (c) and (d) show the two new $\gamma$-ray transitions in the $^{52}$Ca spectra obtained in coincidence with the closest 1427-keV transition and 3150-keV transition, respectively.
    }
    \label{fig:Gamma}
\end{figure}

\begin{figure}[!ht]
    \centering
    \includegraphics[width=\linewidth]{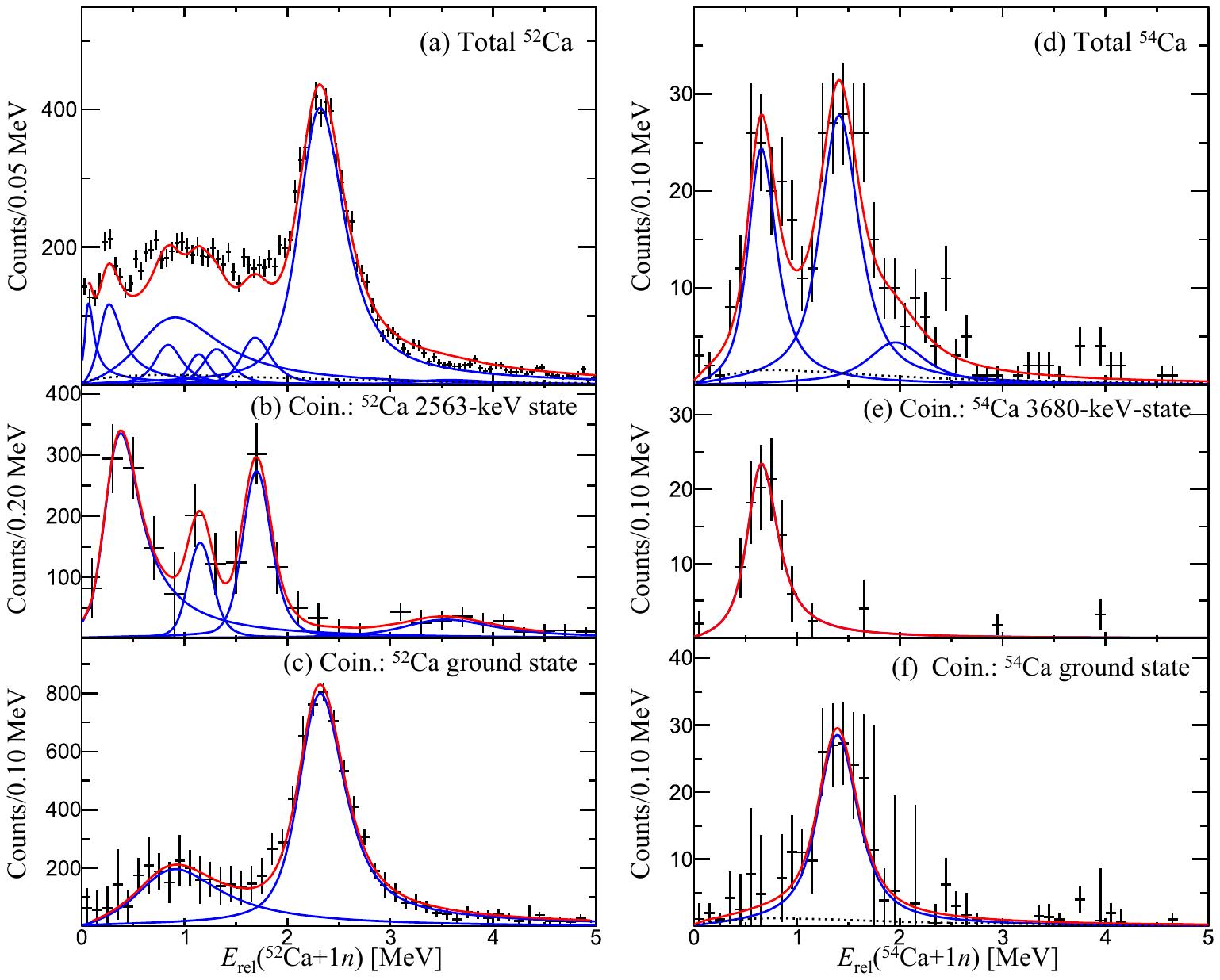}
    \caption{
    (a) Relative energy spectrum of $^{54}\text{Ca}(p,pn)^{53}\text{Ca}\rightarrow$ $^{52}\text{Ca}+1n$; Exclusive relative energy spectra in coincidence with the 2563-keV state (b) and the ground state (c) in \ts{52}Ca. (d) Relative energy spectrum of $^{56}\text{Ca}(p,pn)^{55}\text{Ca}\rightarrow$ $^{54}\text{Ca}+1n$; Exclusive relative energy spectra in coincidence with the 3680-keV state in \ts{54}Ca (e), and ground state (f). The overall fit function (red line) comprises the sum of the simulated responses of NeuLAND + NEBULA and beamline detectors (blue lines), and a non-resonant background (black long-dashed line).}
    \label{fig:Erel}
\end{figure}

Unbound states of $^{53,55}$Ca populated from 1$n$ knockout reactions instantaneously decay into systems of neutrons plus calcium fragments.
The fragments may be in excited states, ensuing $\gamma$-ray de-excitation. Accordingly,
properties of unbound states were reconstructed by invariant-mass spectroscopy.
The excitation energy of an unbound state is written as 
\begin{align}
E_{ex} = E_{rel} + S_{n}+ (E_{\gamma}),
\end{align}
where the $E_{rel}$ refers to the relative energy of the unbound system reconstructed from the momenta of the fragment and neutron.
$S_{n}$ represents the one neutron separation energy, while
$E_{\gamma}$ is the summed $\gamma$-ray energy following the neutron emission, if any.

\paragraph{Doppler-shift corrected $\gamma$-ray spectra}
As exclusive final state and momentum distributions require knowledge
on coincident $\gamma$-ray decay, they are presented first.
Figure~\ref{fig:Gamma} displays the Doppler-corrected $\gamma$-ray spectra of $^{52,54}\text{Ca}$ fragments.
For $^{52}$Ca, peaks at 2563, 1427, 1961, 3150, and 3388\,keV are found, consistent with the transitions reported from a $\beta$-decay study~\cite{perrot:2006:PRC}. 
In addition, two new $\gamma$-ray transitions are observed at 553(9) and 950(17)\,keV in this work, resulting in tentative level placements at 4543(10) and 4100(17)\,keV, according to the $\gamma$-$\gamma$ coincidence analysis, shown in Fig.~\ref{fig:Gamma}(c, d).
The population percentages for 2563, 3150, 3990, 4100, 4543, and 5951 keV states in $^{52}$Ca, extracted from Fig. \ref{fig:Gamma}(a), are as follows: 51.0(28), 19.6(23), 0.0(46), 7.1(13), 8.3(10), and 14.0(20) $\%$, respectively.

For $^{54}$Ca, $\gamma$-ray energies at 459(12), 1187(23), 1667(41), and 2052(26) keV were extracted from Fig.~\ref{fig:Gamma}(b), consistent with Refs.~\cite{steppenbeck:2013:NATURE,Browne:2021:PRL}.
Considering the error bars of the present work, the obtained energies of $^{54}$Ca have been adopted from Ref.~\cite{Browne:2021:PRL} for further analysis.
The $\gamma$-ray spectra were fitted with a double-exponential background and simulated DALI2$^+$ response functions.
In Fig.\ref{fig:Gamma}(b), the $\gamma$-ray spectrum was almost entirely described by  the response function of populating the 3680-keV state on top of exponential background.
The $\gamma$-ray intensity of the 3680-keV state and possibly the 2466-keV state excludes the possibility of a direct feeding to the 2021-keV state.
Consequently, in the decomposition of the $E_{rel}$ spectra of $^{54}\text{Ca}+1n$, feedings to the 2021-keV state are not considered. 

\paragraph{$E_{rel}$ spectra}~The relative energy spectra of the unbound $^{53,55}$Ca system following the $^{54,56}\text{Ca}(p, pn)$ 
reaction are shown in Fig.~\ref{fig:Erel}.
The top panels (a) and (d) show the inclusive $E_{rel}$ spectra of the unbound states, whereas the other panels depict the exclusive $E_{rel}$ spectra in coincidence with final states of $^{52,54}$Ca fragments. 
The $E_{\gamma}$ spectra were analyzed to ascertain the contributions of fragments' final states, in coincidence with a fixed $E_{rel}$-bin, spanning the entire $E_{rel}$ range of [0, 5] MeV.
Panel (b) and (e) represent examples of the exclusive $E_{rel}$ spectra obtained in coincidence with the 2563-keV state of $^{52}$Ca and the 3680-keV state of $^{54}$Ca, respectively. Contributions to the ground states of the $^{52,54}$Ca fragments, shown in Figs.~\ref{fig:Erel}(c) and (f), were extracted by subtraction of all contributions to excited states from the inclusive spectra.

The $E_{rel}$ spectra were fitted with response functions using Breit-Wigner distributions~\cite{lane:1958:RMP}, folded with the detector response obtained from a GEANT4 simulation of the NeuLAND + NEBULA array and beam-line detectors. 
A non-resonant background was considered, for which the shape was obtained from event mixing~\cite{randisi:2014:PRC,leblond:2018:PRL} and an amplitude fitted together with the response functions of the resonances.
The exclusive $E_{rel}$ spectra, i.e., differences when gating on the bound $\gamma$-decaying states of \ts{52,54}Ca, help distinguishing the overlapping peaks and prepare the resonance candidates for the fit of the inclusive $E_{rel}$ spectra.
The exclusive cross-sections to the unbound states of $^{53,55}$Ca were derived from the amplitudes of the Breit-Wigner peaks corrected with the simulated 1$n$-detection efficiency.
Note that $\sim8\%$ of events originating from the inelastic excitation process, as discussed in detail in Refs.~\cite{enciu:2022:PRL,chen:2019:PRL}, were subtracted from the relative energy spectra and momentum distributions.
The fit results are summarized in supplemental Fig.\ref{fig:53Ca-Level} and Fig.~\ref{fig:55Ca-Level}, and Table~1 in Ref.~\cite{supmat:2023:manuscript}.

\begin{figure}[htbp]
    \centering
    \includegraphics[width=\linewidth]{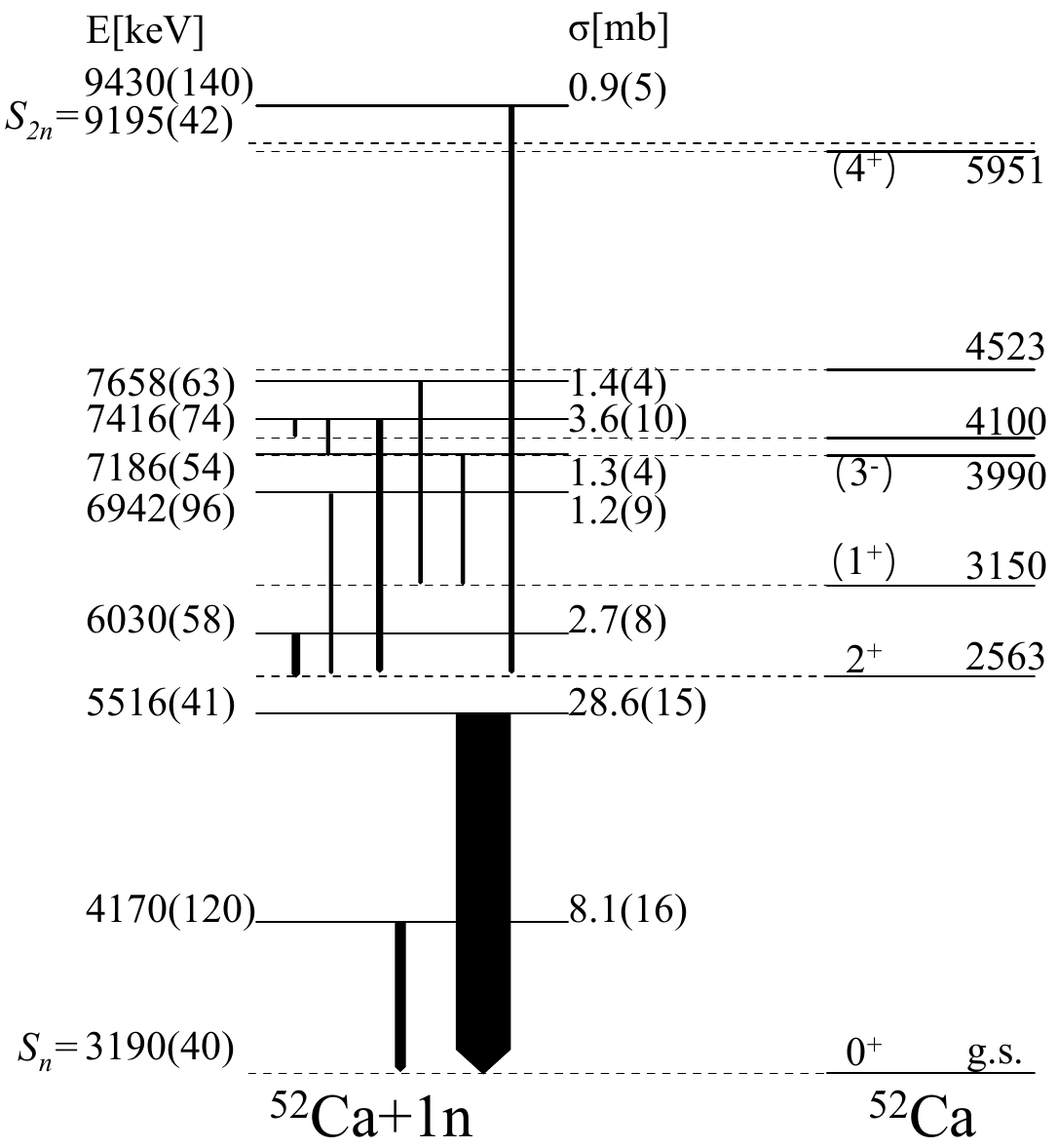}
    \caption{
    Partial level scheme of unbound states in $^{53}$Ca following the $^{54}\text{Ca}(p,pn)$ reaction. The position and width of the unbound states were reconstructed from a detailed analysis of decay neutrons in coincidence with $\gamma$ rays. 
    Arrow widths represent the relative intensity. Bound states of \ts{53}Ca were omitted.}
    \label{fig:53Ca-Level}
\end{figure}

\begin{figure}[htbp]
    \centering
    \includegraphics[width=\linewidth]{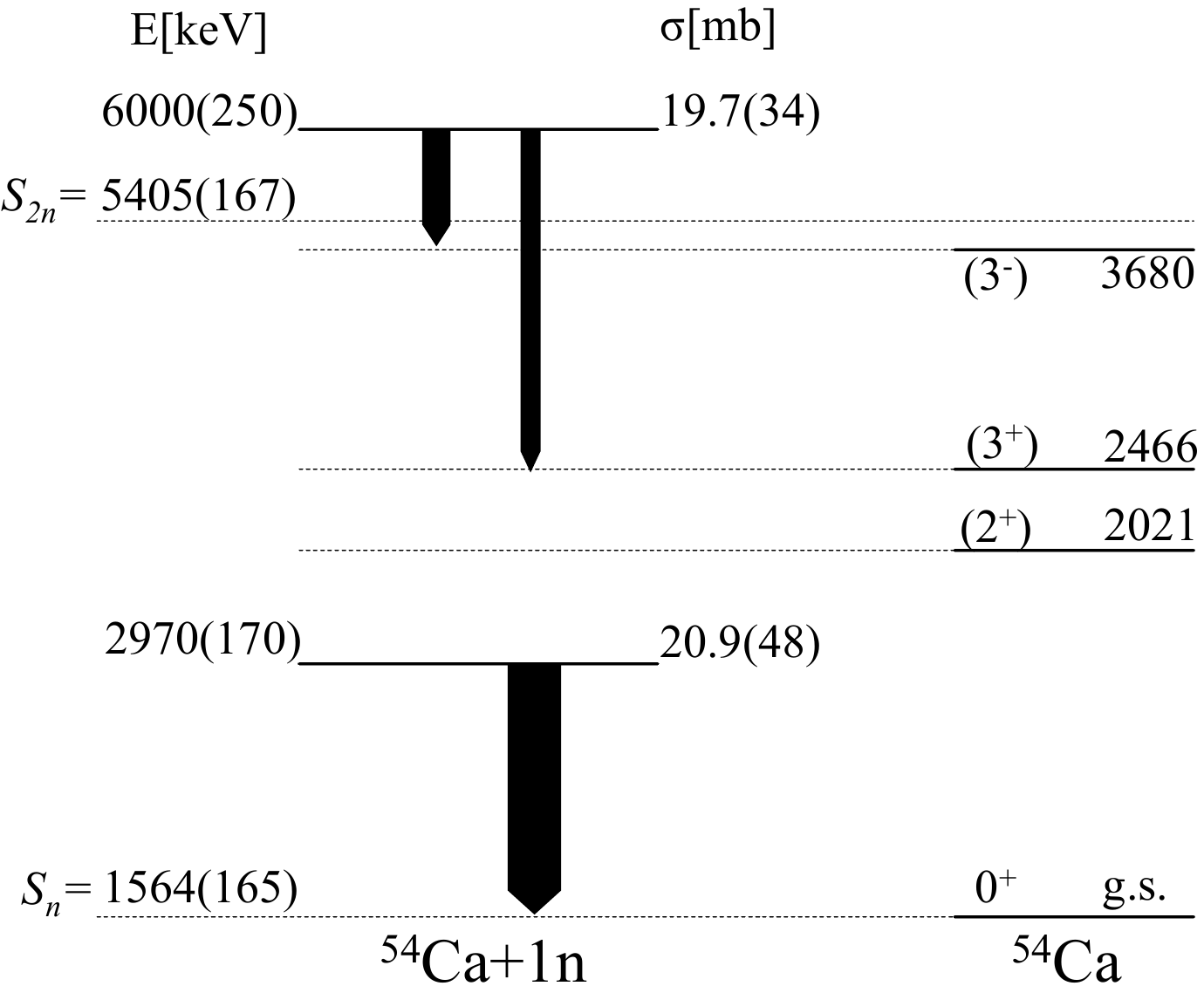}
    \caption{
    Same as Fig.~\ref{fig:53Ca-Level}, but for unbound states in $^{55}$Ca following the $^{56}\text{Ca}(p,pn)$ reaction.}
    \label{fig:55Ca-Level}
\end{figure}

The overall fit of the $^{52}$Ca+$1n$ relative energy spectrum, shown in Fig.~\ref{fig:Erel}(a), is characterized by numerous overlapping peaks, extracted from the exclusive spectra as resonance candidates illustrated in Fig.~\ref{fig:Erel}(b, c) and its supplemental Fig.~1 in Ref.~\cite{supmat:2023:manuscript}. 
Among the observed transitions, approximately 75$\%$ of the measured cross-sections align with the $^{52}$Ca ground state, 13$\%$ with the 2563-keV state, and 12$\%$ with remaining states above the 2$^+_1$ state of $^{52}$Ca.
Of these, two resonances at 980(110) and 2326(10)\,keV were identified to be in coincidence with the $^{52}$Ca ground state [Fig.~\ref{fig:Erel}(c)], corresponding respectively to states at 4170, and 5516\,keV; four resonances at 277(42), 1189(87), 1716(27), and 3680(140)\,keV were in coincidence with the 2563-keV state [Fig.~\ref{fig:Erel}(b)], corresponding respectively to states at 6030, 6942, 7416 and 9430\,keV; further details regarding the remaining four resonances, which were in coincidence with higher excited states, can be found in Ref.~\cite{supmat:2023:manuscript}.
Interestingly, the dominant peak at $2326(10)$\,keV, decaying directly to the ground state of $^{52}$Ca via 1$n$ emission, exhibited a width of $\Gamma = 253(25)$\,keV, approximately aligning with the single-particle width of $\Gamma_{r}^{l=3} = 220$~keV for $f$-wave neutron emission\cite{DOVER:1969:NPA}. 

The $E_{rel}$ spectrum following the $^{56}\text{Ca}(p,pn)^{55}\text{Ca}$ reaction, as shown in Fig.~\ref{fig:Erel}(d), can be clearly described by three resonance peaks added on a non-resonant background. 
Due to low statistics, weaker states with low cross sections could not be effectively observed in this data.
A resonance at 1404(41)\,keV was observed in coincidence with the ground state of $^{54}$Ca, placed at a state of 2970\,keV.
Resonance peaks at 2089(80)\,keV and 664(34)\,keV decay to the 2466-keV state and 3680-keV state of $^{54}$Ca, respectively.
Given the uncertainties, the latter two resonances can be treated as decays from a state at an energy of 6000(250)\,keV.
This state is above the two neutron separation energy of $S_{2n}=5405(167)$\,keV \cite{michimasa:2018:PRL,wang:2021:CPC}, which can also decay by 2$n$ emission. Accordingly, its cross-section in Table~1~\cite{supmat:2023:manuscript} is considered as the lower limit. 
The limited statistics of the $^{53}$Ca$+2n$ channel prevented a detailed analysis of this state. 
Hence, the upper cross-section limit for this state was estimated in the $E_{rel}$ range of [0-1] MeV, considering larger error bars from limited statistics and interference from a resonant-like structure beyond 1 MeV, and subsequently corrected with 2$n$ detection efficiency.
These cross-sections are provided in Table~\ref{tab:C2S} and used for extracting the spectroscopic factors $C^2S$.

\begin{figure}[!ht]
    \centering
    \includegraphics[width=\linewidth]{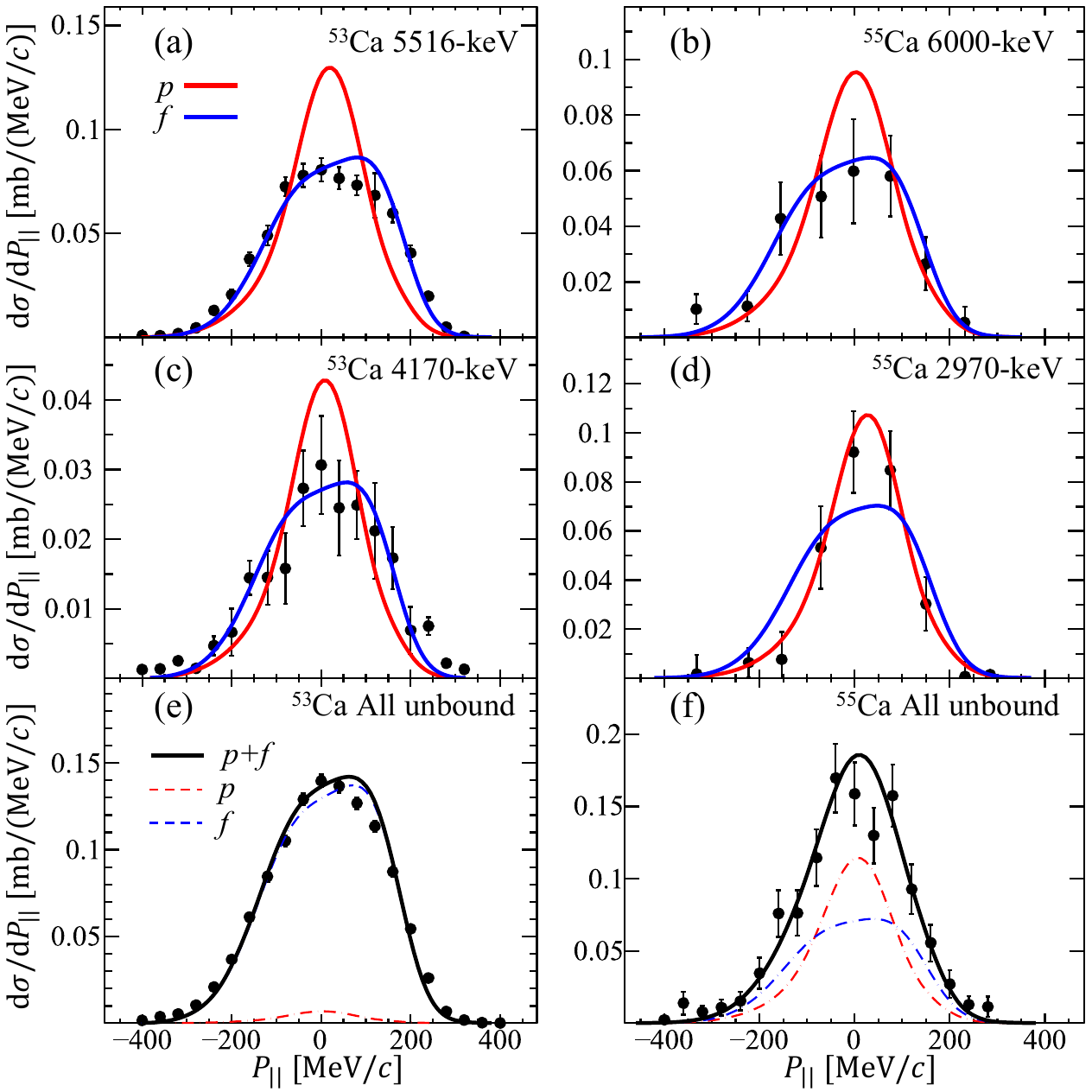}
    \caption{
    Exclusive parallel momentum distributions for (a) 5516-keV and (c) 4170-keV states of $^{53}\text{Ca}$, (b) 6000-keV and (d) 2970-keV states of $^{55}\text{Ca}$. Solid curves are the calculated DWIA predictions assuming an 1$n$ knockout from $p$ and $f$ orbitals and have been normalized to experimental cross-sections. Panel (e) and (f) display PMDs for all the unbound states of $^{53}\text{Ca}$ and $^{55}\text{Ca}$, respectively, compared with a linear fit (black, solid lines) between p-curve (red, dashed) and f-curve (blue, dashed). Error bars represent statistical errors.
    }
    \label{fig:PMD}
\end{figure}

\paragraph{Momentum distributions}~
The momentum distributions of the reaction products were reconstructed using the beam and fragment velocities at the reaction vertex and the scattering angles calculated from the position measurements. 
Exclusive parallel momentum distributions (PMDs) for unbound states of $^{53}$Ca and $^{55}$Ca are presented in Fig.~\ref{fig:PMD}, extracted by fitting the $E_{rel}$ spectra within a fixed PMD-bin, allowing to separate the contributions of the overlapping resonances.
For PMDs, a resolution of 40 MeV/c~$(\sigma)$ was evaluated from the unreacted $^{54,56}$Ca beams.

The orbital angular momentum $l$ of the removed neutron can be deduced from the momentum distribution in neutron knockout reactions.
Theoretical predictions for single-particle cross-sections and momentum distributions of neutron removal from $l=1$ and 3
orbitals were calculated using the distorted wave impulse approximation (DWIA) model~\cite{ogata:2015:PRC,wakasa:2017:PPNP} and then compared with experimental results. 
DWIA calculations have been successfully applied in the quasi-free scattering studies for neutron-rich nuclei in this region~\cite{enciu:2022:PRL,chen:2019:PRL,Browne:2021:PRL,chen:2023:PLB,sun:2020:PLB,linh:2021:PRC}. 
The calculated PMDs have been folded with the experimental resolution, also taking into account the energy spread induced by the thick target.

The PMD of the 5516-keV state in $^{53}$Ca [Fig.~\ref{fig:PMD}(a)] was well reproduced by the $f$-orbital distribution from the DWIA calculation, providing evidence for the $l=3$ assignment of this state. In addition, fitting of the PMD of the 4170-keV state [Fig.~\ref{fig:PMD}(c)] yielded a reduced $\chi^2$ of 0.5 for the $f$-curve, and 2.6 for the $p$-curve, favoring the $l=3$ assignment of neutron removal from the $f$-orbital.  
The 6000-keV state in $^{55}$Ca was tentatively assigned to $l=3$. The PMD of this state, shown in Fig.~\ref{fig:PMD}(b), exhibits a wide distribution and is better fitted with the $f$-curve, yielding a reduced $\chi^2$ value of 0.6. In contrast, the fit using a $p$-curve yields a higher reduced $\chi^2$ of 1.6, 
suggesting a preference for the $l=3$ assignment.
The PMD of the 2970-keV state of $^{55}$Ca [Fig.~\ref{fig:PMD}(d)] was consistent with a $p$-wave neutron knockout, supporting the $l=1$ assignment of this state.
In supplemental Fig.~4~\cite{supmat:2023:manuscript}, the PMD fitting indicates a preference of $l=3$ for the 6030-keV state and $l=1$ for the 7416-keV and 7658-keV states.
Due to limited statistics, other exclusive PMDs did not yield meaningful results. 
More details can be found in Ref.~\cite{supmat:2023:manuscript}.

The PMDs for the entire $E_{rel}$ strength indicated a dominance of $f$-wave neutron removal. Fig.~\ref{fig:PMD}(e) and (f) demonstrate linear fits incorporating the $p$-wave and $f$-wave components. The best fit with relative contributions of two components allows for determining the cross-sections for a neutron knockout from the $l=1$ and 3 orbitals.
The inclusive PMDs of all the unbound states of $^{53}$Ca can be effectively described by the $f$-wave curve. Specifically, the extracted cross-sections of 1.6(12)\,mb and 47.8(29)\,mb correspond to the $p$-wave and $f$-wave components, respectively.
Compared to the total cross-section of 39.4(23)\,mb for the $7/2^-$ states listed in Table~\ref{tab:C2S}, the extracted $f$-components fall outside the accepted error margins.
This departure primarily results from the underestimation of the $p$-components, attributed to the inadequate fit in the PMDs of [0, 120] MeV regions. 
Additionally, the relatively high $S_{2n}$ allows the potential contribution from knockout processes within the $sd$ shell, which is beyond the scope of this study.
In Fig.~\ref{fig:PMD}(f), the total cross-section extracted for all the unbound states of $^{55}$Ca amounts to 46.7(44) mb, in accordance with the cross-sections of the $f$ and $p$ components as listed in Table~\ref{tab:C2S},  accounting for the non-negligible background demonstrated in Fig~\ref{fig:Erel}(d).
In the fitting processes, the ratio between the $p$-wave and $f$-wave components has been constrained using the cross-sections in Table ~\ref{tab:C2S}, resulting in a fit with a reduced $\chi^2$ of 1.05. 
Allowing this ratio to vary freely improved the fit to a reduced $\chi^2$ of 0.84, albeit with the $p$-component contribution decreasing by approximately 50$\%$.
Importantly, the extracted cross-sections remain consistent with Table~\ref{tab:C2S}'s results and fall within the associated error bars. 

\paragraph{Spectroscopic factors}~Spectroscopic factors $C^{2}S$ were obtained by comparing the measured cross-sections to single particle cross-sections $\sigma_{sp}$ of the DWIA calculations, shown in Table~\ref{tab:C2S}.
A systematic uncertainty of 15$\%$ was considered for the calculated  $\sigma_{sp}$~\cite{wakasa:2017:PPNP}. 
Quenching factors were not applied due to the proximity of $C^2S$ to the independent particle expectation of $2j+1$ value and the additional uncertainties associated with their application, given the debated isospin dependence\cite{Aumann:2021:PPNP} and contradictory results using hydrogen versus traditional carbon or beryllium targets\cite{atar:2018:PRL,wamers:2023:EPJA}.

A cross-section of 28.6(15)\,mb was determined for the 5516-keV state in $^{53}$Ca, yielding a $C^2S$ of 6.5(3).
This observation suggests the presence of a nearly pure $\nu 0f_{7/2}$ neutron-hole state, aligning with an extreme single-particle description.
Two neighboring states at 4170\,keV and 6030\,keV are tentatively assigned to a spin-parity of $7/2^-$, based on the exclusive PMD fitting.
Their cross-sections are 8.1(16)\,mb and 2.7(8)\,mb, resulting in $C^2S$ values of 1.8(3) and 0.6(2), respectively.
In Ref.~\cite{chen:2019:PRL}, the ground state of $^{53}$Ca was tentatively assigned to a spin-parity of 1/2$^-$, associated with $\nu p_{1/2}$ neutron knockout with a $C^2S$ of 2.2(2); and the 2220-keV state is tentatively assigned to a spin-parity of 3/2$^-$ associated with $\nu p_{3/2}$ neutron knockout with a $C^2S$ of 3.1(2).
Concerning the 7416-keV and 7658-keV states, favoring the $l=1$ in the PMD fitting, a plausible assignment of a spin-parity of $3/2^−$ is proposed, yielding $C^2S$ values of 0.9(3) and 0.3(1), respectively.
However, the unfavorable orbital assignment can not be entirely excluded due to the error bars. 
Spectroscopic factors could not be determined for other unbound states in \ts{53}Ca, as their respective PMD and accordingly $l$-value could not be 
determined. 


For $^{55}$Ca, only two bound states have been identified from one-neutron knockout: The ground state was tentatively assigned to a spin-parity of $5/2^-$ with a $C^2S$ of $2.0^{+1.6}_{-0.9}$, and the 673-keV state tentatively assigned to a spin-parity of $1/2^-$ with a $C^2S$ of $3.0^{+1.0}_{-0.6}$~\cite{koiwai:2022:PLB}. 
The present PMDs favor a spin parity assignment of $3/2^-$ for the 2970-keV state. Accordingly,
the calculated single particle cross section for an $l=1$ states is 5.72\,mb, resulting in an
$C^2S$ extracted to be 3.7(8). Similarly, the 6000-keV state can be attributed to spin parity
$7/2^-$ and a $C^2S$ estimated to be 4.8-5.9(9). Here, the lower limit originates from the 1$n$ emission data, while an upper limit was estimated by taking into account the 2$n$ emission.



\begin{table*}
\centering
\caption{
Identified unbound neutron-hole states populated from $^{54,56}\text{Ca}(p, pn)$ reactions. Excitation energies, cross sections, and spectroscopic factor ($C^2S$) are compared with theoretical values from shell model calculations using the GXPF1Bs interaction and \textit{ab initio} VS-IMSRG calculations employing the chiral NN+3N interactions [1.8/2.0 (EM) and $\Delta$N$^2$LO$_\text{GO}$].
Spin-parity assignments are based on systematics and theoretical calculations. The experimental $C^2S$ were deduced by dividing the measured cross sections by the calculated single particle cross-section from the DWIA framework. Level energies are given in keV and cross-sections in mb.
}
\label{tab:C2S}
\small
\begin{tabular}{cccccp{0.03\textwidth}p{0.03\textwidth}p{0.04\textwidth}p{0.04\textwidth}p{0.03\textwidth}p{0.03\textwidth}p{0.04\textwidth}p{0.04\textwidth}p{0.03\textwidth}p{0.03\textwidth}}
\toprule
& & & &DWIA &\multicolumn{2}{c}{GXPF1Bs}&\multicolumn{2}{c}{$pf$1.8/2.0(EM)}&\multicolumn{2}{c}{$pf$$\Delta$N$^2$LO$_\text{GO}$}&\multicolumn{2}{c}{$pfg$1.8/2.0(EM)}&\multicolumn{2}{c}{$pfg$$\Delta$N$^2$LO$_\text{GO}$}\\
\cmidrule(lr){6-7} \cmidrule(lr){8-9} \cmidrule(lr){10-11} \cmidrule(lr){12-13} \cmidrule(lr){14-15}
E&$\sigma^{exp}_{-1n}$&$C^2S_{exp}$&$J^{\pi}$&$\sigma^{sp}$&E&$C^2S$&E&$C^2S$&E&$C^2S$&E&$C^2S$&E&$C^2S$\\ 
\midrule
\multicolumn{15}{c}{$^{54}\text{Ca}(p,pn)^{53}\text{Ca}$$\rightarrow$$^{52}\text{Ca}+1n$}\\
4170(120) & 8.1(16) & $1.8(3)(3)^{1}$ & $7/2^-$ & 4.59 & 4985$^4$ & 0.4$^4$&7042$^4$& 0.4$^4$& 7293$^4$&0.2$^4$ & 7216$^4$&0.1$^4$&6387$^4$& 0.6$^4$\\
5516(41) & 28.6(15) & $6.5(3)(10)^{1}$ & $7/2^-$ & 4.38 & 5225 & 7.1&6756& 7.3& 7053&7.3 & 6454&7.5&6633& 6.8\\
6030(58) & 2.7(8) & $0.6(2)(1)^{1}$ & $7/2^-$ & 4.31 & & & & & & & & & & \\
\midrule
\multicolumn{15}{c}{$^{56}\text{Ca}(p,pn)^{55}\text{Ca}$$\rightarrow$$^{54}\text{Ca}+1n$}\\ 
2970(170) & 20.9(48)& $3.7(8)(5)^{1}$ & $3/2^-$ & 5.72 & 3152 & 3.3& 3975& 3.7& 3007& 3.7 & 3728&3.6&2805&3.6\\
6000(250) & 19.7(34)$^2$ & $4.8(8)(7)^{1}$ & $7/2^-$ & 4.10 & 5789 & 3.8& 8023& 6.8&7681& 7.1&7344&4.6&7000&6.2\\
6000(250) & 24.2(36)$^3$ & $5.9(9)(8)^{1}$ & $7/2^-$ & 4.10 & 5927$^4$& 1.5$^4$&9008$^4$ & 0.6$^4$& 8353$^4$& 0.3$^4$&7431$^4$&2.1$^4$&6972$^4$&1.0$^4$\\
\bottomrule
\end{tabular}
\begin{minipage}[t]{\linewidth}
\footnotesize
1. The $1^{st}$ and $2^{nd}$ parentheses represent the statistical error and the systematic error originating from the uncertainty of $\sigma_{sp}$\cite{wakasa:2017:PPNP}, respectively.\\
2. Lower limits obtained from the 1n emission data.\\
3. Upper limits estimated from the 2n emission data.\\
4. The second largest $C^2S$ value in the calculations.
\end{minipage}
\end{table*}

\section{Discussion}

The observation of $0f_{7/2}$ neutron-hole states in the quasi-free scattering reaction \ts{54,56}Ca$(p,pn)$\ts{53,55}Ca reveals a shell gap of approximately 3\,MeV between the $0f_{7/2}$ and $1p_{3/2}$ orbitals, with large spectroscopic factors indicating the robustness of the $N=28$ shell closure.
Hence, a large neutron excess in the Ca isotopes seems not to result in $N=28$ shell gap degradation.
Neutrons in the deeply bound $0f_{7/2}$ shell are arranged  
as in the extreme single particle picture, providing further evidence for the 
magicity of $N=32$ and $N=34$.
In contrast, the results for $^{55}$Ca demonstrate more pronounced fragmentation in the deeply bound $\nu 0f_{7/2}$ orbital.
These findings on location and occupation of the deeply bound orbitals offer an unprecedented opportunity to test 
state-of-the art theories up to high excitation energies beyond the neutron separation energies.

\begin{figure*}[!ht]
    \centering
    \includegraphics[width=\linewidth]{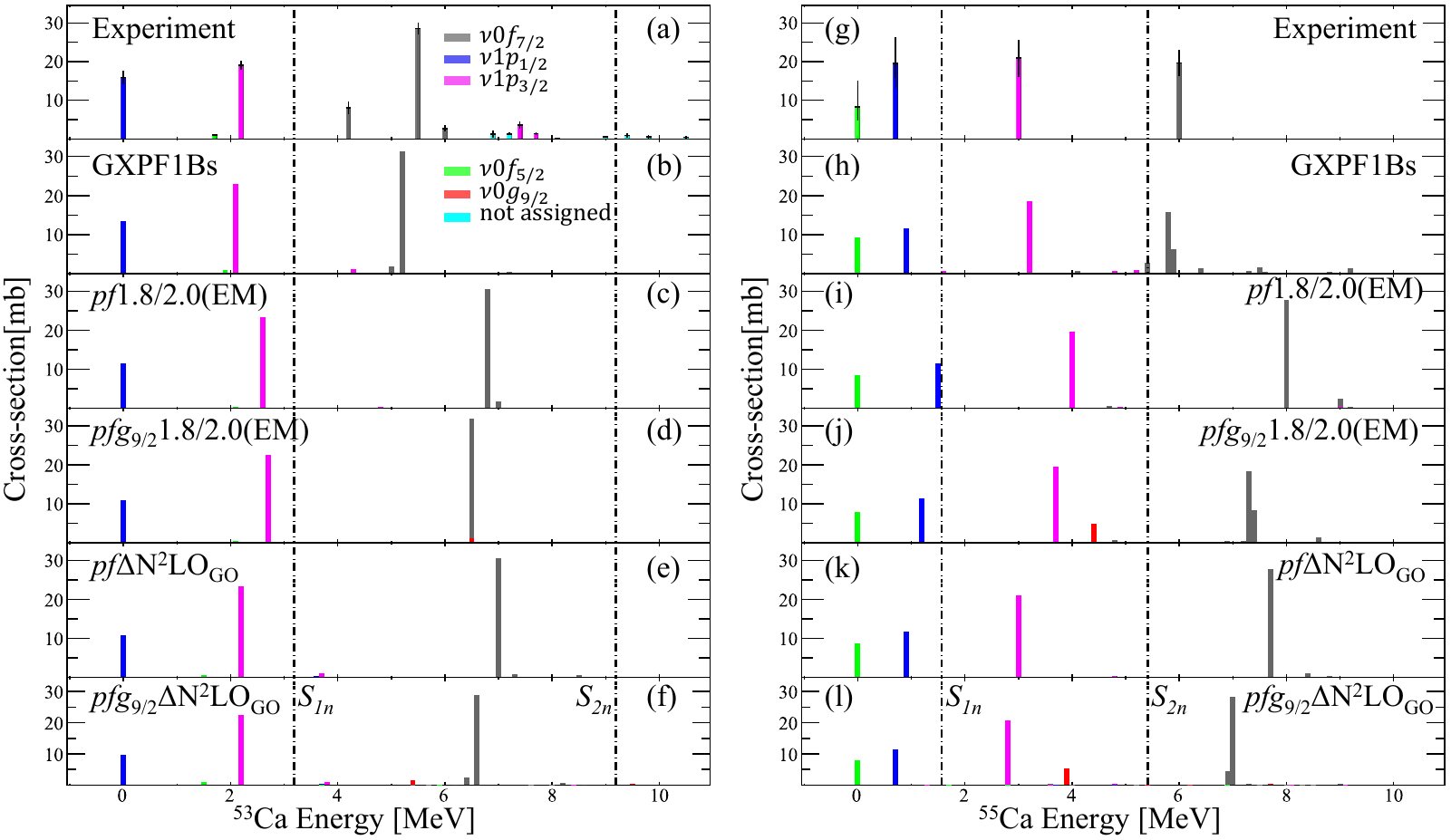}
    \caption{Panel (a) and (g) depict the measured cross-sections for $^{53}$Ca and $^{55}$Ca as a function of excitation energy. The cross-sections to bound states below $S_{1n}$ were deduced from $\gamma$-ray spectroscopy measurements\cite{chen:2019:PRL,koiwai:2022:PLB}. The cross-sections to the unbound states (this work) were obtained from invariant mass spectroscopy. Identified neutron-hole states are colored accordingly to the orbital of the removed neutron.
    The other states with no conclusive PMDs are labeled "not assigned".
    Panel(b-f) are the theoretical predictions of energies and cross-sections. Predictions associated with the $\nu 0g_{9/2}$ orbital are scaled by a factor of 5.
    }
    \label{fig:C2S}
\end{figure*}

The experimental cross-sections are confronted with theoretical predictions obtained by combining the $\sigma_{sp}$ with the $C^2S$ values from the shell model or \textit{ab-initio} calculations, shown in Table~\ref{tab:C2S} and Fig.~\ref{fig:C2S}. 
Shell model calculations were performed using the latest GXPF1Bs Hamiltonian~\cite{chen:2019:PRL} from the GXPF1 family of effective nucleon-nucleon interactions~\cite{honma:2005:EPJA} in the $pf$ model space. 
These calculations have also been effectively employed to interpret the bound states results for \ts{54}Ca$(p,pn)$\ts{53}Ca~\cite{chen:2019:PRL} and \ts{56}Ca$(p,pn)$\ts{55}Ca~\cite{koiwai:2022:PLB} reactions.
The VS-IMSRG~\cite{stroberg:2019:ARNPS,tsukiyama:2012:PRC,hergert:2016:PR,stroberg:2017:PRL} calculations were 
carried out employing two different sets of chiral NN+3N interactions: 1.8/2.0 (EM)~\cite{hebeler:2011:PRC,Simonis:2017:PRC} and 
$\Delta$N$^2$LO$_\text{GO}$~\cite{jiang:2020:PRC}, derived from chiral effective field theory. 
The former interaction has proven its effectiveness in numerous studies~\cite{Morris:2018:PRL,taniuchi:2019:nature,stroberg:2021:PRL}, while the latter interaction, optimized using properties of A$\leq$4 nuclei and nuclear matter, exhibits remarkable capability to reproduce binding energies and charge radii from $A=16$ to $A = 132$~\cite{jiang:2020:PRC}.
The chosen model space used an $^{40}$Ca core with valence neutrons in the $pf$ or $pfg_{9/2}$ shell using a multi-shell variant of the VS-IMSRG~\cite{Miyagi:2021:PRC}.

As shown in Fig.~\ref{fig:C2S}, the experimental cross-sections in panel (a,g) are in good agreement with theoretical predictions in panels (b-f,h-l) which are derived from both SM and \textit{ab-initio} calculations. 
The SM predictions using the effective GXPF1Bs Hamiltonian are remarkably consistent with the experimental results for both bound states~\cite{chen:2019:PRL,koiwai:2022:PLB} and the unbound states observed 
in this work. Calculated energy and $C^2S$ values strongly reinforce the experimental spin assignments.

As for the VS-IMSRG calculations, the predicted cross-sections of major neutron-hole states are in line with the data.
However, the position of the $0f_{7/2}$ hole-state is predicted to be approximately 1\,MeV higher for \ts{53}Ca, and between 1 and 
2 MeV higher for \ts{55}Ca. More generally, utilization of the 1.8/2.0 (EM) interaction leads to higher energies for the $1p_{1/2}$, $1p_{3/2}$ and $0f_{7/2}$ neutron-hole states.
Conversely, using the $\Delta$N$^2$LO$_\text{GO}$ interaction leads to predicted energies for the $1p_{1/2}$ and $1p_{3/2}$ neutron-hole states within 200-keV difference from the experimental data, but results in a shell gap larger than 4\,MeV between the $0p_{3/2}$ and $0f_{7/2}$ orbitals. 
Another interesting observation is that model space expansion from the $pf$ shell to the $pf+g_{9/2}$ orbitals leads to an improved agreement between the predictions and experimental results. Specifically, the location of the $0f_{7/2}$ hole state is lowered by about 500 keV for \ts{53}Ca and 1 MeV for \ts{55}Ca.
Furthermore, the $0f_{7/2}$ strength fragments into two major states in panel (j) and (l) for the isotope $^{55}$Ca, while a single major hole state is evident in panel (d) and (f) for $^{53}$Ca, and thus consistent with the GXPF1Bs Hamiltonian. As the two states in \ts{55}Ca are predicted close in energy, they may have remained unobserved due to their width and/or the experimental resolution. 
Further improvements are needed in the VS-IMSRG framework, including the consideration of continuum effects on unbound states\cite{hagen:2012:PRL}.
It is noteworthy that these effects are almost negligible for the $7/2^{−}$ states due to its significant centrifugal barrier. 
The GXPF1Bs calculations, based on an effective interaction, do not explicitly include continuum effects.
Ref.~\cite{kondo:2023:Nature}  demonstrates that there is no significant difference between effective interactions and those with explicit continuum coupling, such as the Gamow shell model.

\begin{figure*}[!ht]
    \centering
    \includegraphics[width=\linewidth]{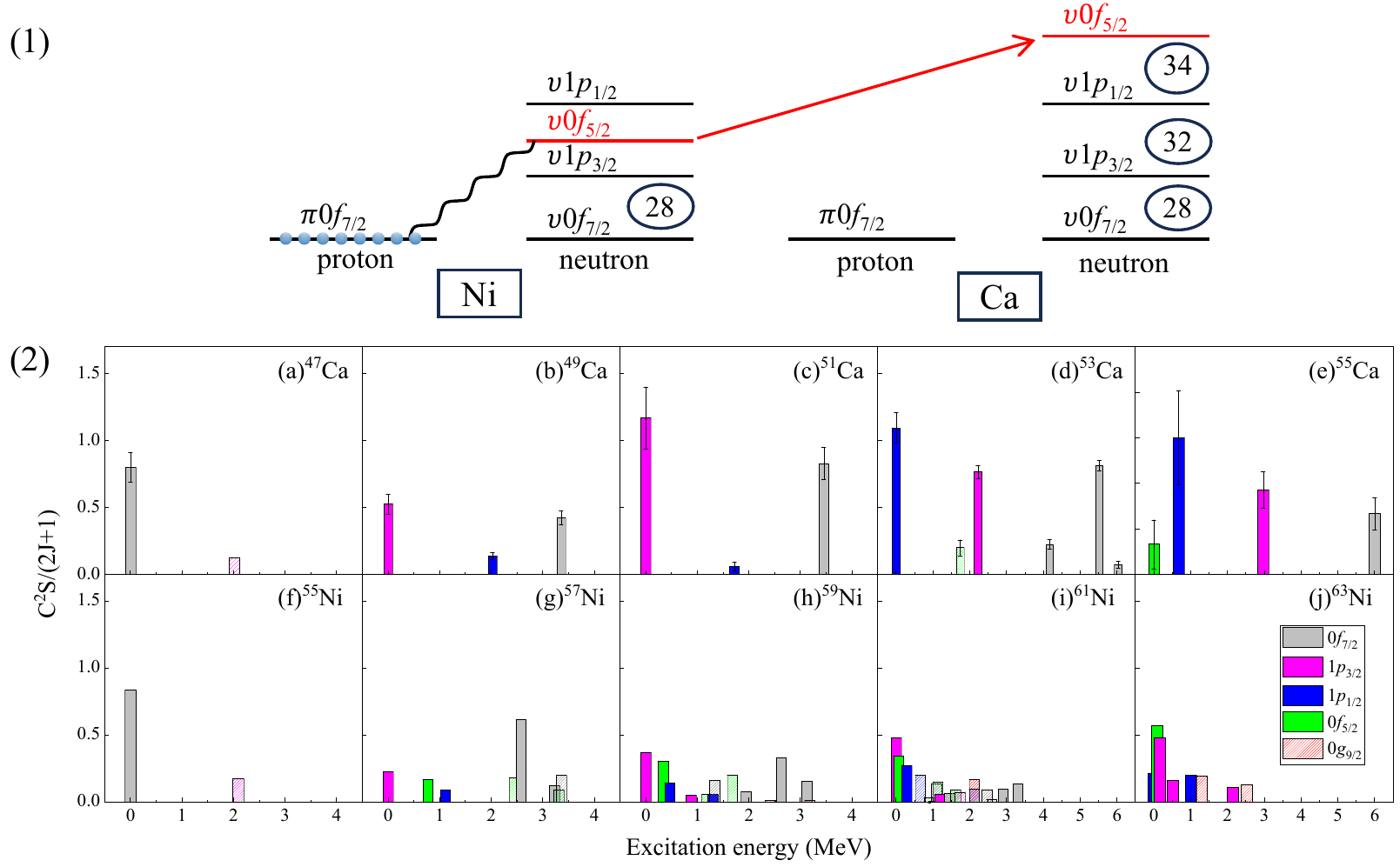}
    \caption{(1) Schematic illustration of shell evolution from Ni to Ca for neutron $pf$ orbitals. The wavy line implies the attractive interaction between the proton $0f_{7/2}$ and the neutron $0f_{5/2}$ orbits. Adopted from Ref~\cite{otsuka:2016:JPG}. (2) Normalized spectroscopic factors of neutron-hole states as a function of excitation energy in the Ca($Z=20$) and Ni($Z=28$) isotopes, populated from neutron removal reaction. Data of Ca isotopes are taken from neutron knockout reactions~\cite{crawford:2017:PRC, enciu:2022:PRL,chen:2019:PRL,koiwai:2022:PLB} and the present work (unbound states); Data of Ni isotopes are taken from $(p,d)$ or $(^{3}\text{He},\alpha)$ transfer reactions~\cite{sanetullaev:2014:PLB,schiffer:2013:PRC}. Values where $C^{2}S/(2J+1) < 0.05$ have been multiplied by a factor of 5 and are displayed with hatching in the color corresponding to the respective orbital.
    }
    \label{fig:Ca-Ni}
\end{figure*}

From the systematics of the $C^2S$ in the Ca isotopes, shown in Fig.~\ref{fig:Ca-Ni}(b), the $0f_{7/2}$ strength maintains a nearly pure neutron-hole character in reaction residues populated by neutron knockout from the doubly magic nuclei $^{48,52,54}$Ca.
These results are particularly intriguing, as it provides further evidence for the $N=32,34$ sub-shell closures.
Stronger fragmentation of the $0f_{7/2}$ neutron hole state strength in $^{55}$Ca is expected, given the difficulty in maintaining a pure single-hole character at approximately 6\,MeV. The experiment did not observe significant population to higher-lying $0f_{7/2}$ hole states, 
possibly due to their low cross sections or decay via 2$n$ emission channel with limited statistics.
In contrast, previous results for \ts{49}Ca, finding a considerable reduction of $\nu 0f_{7/2}$ strength and a spectroscopic factor of 3.4(4) for the $7/2^{-}$ state at 3357 keV~\cite{crawford:2017:PRC}, demonstrate a slight deviation from current experimental findings.
These results could not be reproduced by any calculations, and suggest a larger degree of fragmentation compared to $^{55}$Ca. 
It is also stressed that for \textit{ab-initio} calculations, 
a considerable fragmentation occurs for \ts{55}Ca only with the inclusion of the $0g_{9/2}$ orbital. 
In contrast, for the GXPF1 family of interactions this fragmentation is seen already with a model space restricted to the $pf$-shell. 

Figure~\ref{fig:Ca-Ni}(1) illustrates the evolution of the neutron $pf$ orbitals from nickel to calcium isotopes.
In Ca isotopes where $\pi 0f_{7/2}$ is unoccupied, the vanishing $j_{>}-j_{<}$ proton-neutron coupling shifts the $\nu 0f_{5/2}$ orbital above the $\nu 1p_{1/2}$ orbital, leaving a gap at $N = 32$ and forming another gap at $N = 34$~\cite{otsuka:2016:JPG,otsuka:2020:RMP}.
Figure~\ref{fig:Ca-Ni}(2) presents a comparative analysis of the spectroscopic strengths of neutron-hole states in the calcium and nickel isotopes, populated from one neutron removal reaction.
This analysis establishes the trend in spectroscopic strength distribution and allows us to examine the impact of the fully occupied or empty $\pi 0f_{7/2}$ shell on the neutron $pf$-shell configuration.
Its occupation has significant consequences on the strength distribution of deeply bound single-particle states in the $pf$ orbitals.
For the $N$ = 27 isotones, the ground states of $^{47}$Ca and $^{55}$Ni exhibit a similar pattern, featuring a nearly pure neutron-hole structure that aligns with the independent particle shell picture. Already for $N$ = 29 isotones, differences between the calcium and nickel emerge: While the ground state of $^{57}$Ni shows half the $C^2S$ compared to the ground state of $^{49}$Ca, more strength associated with the $\nu 0f_{5/2}$ orbital is found for the other low-lying states.
For neutron numbers $N=$31 and larger, the calcium isotopes keep the relatively large spacing between the four orbitals in the neutron 
$pf$-shell, and the neutron-hole states maintain a nearly pure hole nature with limited fragmentation.
In contrast, the occupation of the $\pi 0f_{7/2}$ orbital in the nickel isotopes leads to level compression between $1p_{1/2}$, $1p_{3/2}$, and $0f_{5/2}$ orbitals (<1\,MeV), promoting cross-shell excitations, and further strength fragmentation for the $pf$ orbitals.

\section{Conclusion}

In summary, unbound states of $^{53}$Ca and $^{55}$Ca were measured for the first time using the quasi-free scattering
reactions $^{54,56}$Ca($p,pn$), respectively, providing intriguing results on the shell evolution in the Ca isotopes 
near $N=34$.
Resonance energies, exclusive cross sections, and momentum distributions have been extracted, resulting in the observation of eight unbound states in $^{53}$Ca and two unbound states in $^{55}$Ca.
The 5516-keV state in $^{53}$Ca was found to have a momentum distribution of $l=3$ and was tentatively assigned to have a spin-parity of $7/2^-$ with a $C^2S$ of 6.5(3), indicating a nearly pure neutron-hole state.
For $^{55}$Ca, the 6000-keV state was tentatively assigned a spin-parity of $7/2^-$ with a $C^2S$ of 4.8-5.9(9), while the 2970-keV state was tentatively assigned a spin-parity of $3/2^-$ with a $C^2S$ of 3.7(8).
Our findings are supported by theoretical predictions using the DWIA framework with calculations either utilizing the GXPF1Bs Hamiltonian or the VS-IMSRG method employing the 1.8/2.0(EM) and $\Delta$N$^2$LO$_\text{GO}$ interactions.
Notably, calculations using the GXPF1Bs interaction exhibited a remarkable consistency with the experimental data, both in terms of resonance energies and $C^2S$ values.
A shell gap of approximately 3 MeV between the $0f_{7/2}$ and $1p_{3/2}$ orbitals is found in neutron-rich calcium isotopes, which was deduced from the neutron-hole states with high $C^2S$ values, indicating the robustness of the $N=28$ shell closure, even with the deeply bound $\nu 0f_{7/2}$ orbital.
The persistence of the $0f_{7/2}$ single-particle strength up to $^{54}$Ca, along with its fragmentation in $^{56}$Ca, underscores the absence of the possible $0f_{5/2}$ and $0g_{9/2}$ components in the closed-shell nucleus of $^{54}$Ca, providing further evidence for the $N=34$ magicity.

In view that the considerable reduction of $\nu 0f_{7/2}$ strength to the neutron-hole state in $^{49}$Ca was observed in the neutron knockout from $^{50}$Ca on a beryllium target at lower velocities, it is desirable to study the entire calcium isotopic change with a solid and a liquid hydrogen target at the same energies in order to further elucidate differences in the reaction mechanism\cite{Aumann:2013:PRC,panin:2021:EPJA,Pohl:2023:PRL,Crawford:2020:APR,Petri:S467:R3B}.
\section*{Acknowledgments}
We would like to express our gratitude to the RIKEN Nishina Center
accelerator staff for providing the stable and high-intensity primary beam and to the BigRIPS team for operating the secondary beams.
P.J.L. acknowledges the support from China Postdoctoral Science Foundation(Grant No. YJ20210186).
J.L. acknowledges the support from Research Grants Council (RGC) of Hong Kong with grant of Early Career Scheme (ECS-27303915).
S.C. acknowledges the support of the IPA program at RIKEN Nishina Center.
F.B. acknowledges the support of the Special Postdoctoral Researcher
Program.
The VS-IMSRG calculations were performed with support from
NSERC under grants SAPIN-2018-00027 and RGPAS-
2018-522453, the Arthur B. McDonald Canadian Astroparticle Physics Research Institute, and The Digital Research Alliance of Canada.
T.A. acknowledges funding by the Deutsche Forschungsgemeinschaft (DFG, German Research Foundation) - Project-ID 279384907 - SFB 1245.
Y.T. acknowledges the support from JSPS Grant-in-Aid for Scientific Research Grants No. JP21H01114.
D.S. was supported by the National Research, Development and Innovation Fund of Hungary (NKFIH), financed by the project with contract no. TKP2021-NKTA-42 and under the K18 funding scheme with project no. K128947.



\bibliographystyle{elsarticle-num} 
\bibliography{references}

%
%
%
\end{document}